\RequirePackage{fix-cm}
\documentclass[smallextended]{svjour3}       
\smartqed  
\usepackage{graphicx}
\newtheorem{principle}{Principle}
\newtheorem{observation}{Observation}

\begin{document}

\title{Limitations imposed by complementarity
}


\author{F. E. S. Steinhoff         \and
        M. C. de Oliveira 
}


\institute{F. E. S. Steinhoff \at
              Universidade Federal de Mato Grosso, Faculdade de Engenharia, V\'arzea Grande, 78060-900, Brazil\\
              Universit\"at Siegen, Naturwissenschaftlich Technische Fakult\"at, Siegen, D-57068, Germany \\
              \email{steinhofffrank@gmail.com}           
           \and
           M. C. de Oliveira \at
              Universidade Estadual de Campinas, Instituto de F\'isica ``Gleb Wataghin'', Campinas, 13083-970, Brazil
}

\date{Received: date / Accepted: date}

\maketitle

\begin{abstract}
Complementarity is one of the main features of quantum physics that radically departs from classical notions. Here we consider the limitations that this principle imposes due to the unpredictability of measurement outcomes of incompatible observables. 
For two-level systems, it is shown that any preparation violating complementarity enables the preparation of a non-signalling box violating Tsirelson's bound. Moreover, these ``beyond-quantum'' objects could be used to distinguish a plethora of non-orthogonal quantum states and hence enable improved cloning protocols. 
For higher-dimensional systems the main ideas are sketched.
\end{abstract}

\section{Introduction}
\label{intro}

The basic postulates and results of a physical theory are based on principles that are strongly supported by empirical evidence. 
The principle of conservation of energy, for example, is a major pillar in all areas of physics and implies deep limitations on human experience:
it is impossible to construct a perpetual motion machine, or to outperform Carnot's heat engine. General relativity theory is governed by the equivalence principle
and by the bound on the maximal speed of interactions given by the speed of light. These and other celebrated principles are often not only simple to understand, 
but very precisely stated, giving profound intuitions on the laws of nature. 

In quantum theory, there is an ongoing search for one or more physical principles that could explain the bounds on quantum correlations. More precisely, even though quantum systems can surpass classical bounds of Bell-like inequalities, it is known that there are limits to the violations of local realism attained by quantum objects. There exist theoretical constructions 
known as nonlocal boxes which can violate Bell-like inequalities more than quantum systems, without violating the principle of non-signalling or basic probability axioms. There are many different
 proposals of physical principles that try to explain such bounds on quantum correlations \cite{principles1,principles2,principles3,principles4,principles5}, but there is no general consensus on their success \cite{aquantum}. 

For quantum systems, a major law is the principle of complementarity, firstly noticed by Bohr \cite{up1,bohr1}, based on the observation that certainty in 
the measurement of a fixed physical property  
precludes certainty in the measurement of a complementary one. 
In the double-slit experiment, complementarity is quantitatively expressed by the duality relation
\begin{eqnarray}
D^2+V^2\leq 1, \label{dualrel}
\end{eqnarray}
where $D$ is the path distinguishability and $V$ is the fringe visibility, verified by both empirical \cite{dsexp} and theoretical \cite{dstheo1,dstheo2} methods. 
According to Feynman, the double-slit experiment  ``has in it the heart of quantum mechanics; in reality it contains the only mistery" of the theory \cite{feynman}. 
Applications can be found in Wheeler's delayed choice experiment \cite{wheeler}, which culminated in the concept of the quantum eraser \cite{qeraser1,qeraser2,qeraser3,qeraser4,qeraser5,qeraser6,qeraser7}. 
More recently there is a growing interest 
in re-interpreting complementarity \cite{compint,compintexp1,compintexp2,compintexp3,compintexp4,compintexp5} without, however, violating the empirical relation (\ref{dualrel}). 
The purpose of the present contribution is to consider the implications of a hypothetical violation of complementarity. At least one experiment claiming a maximal violation of this relation was proposed by Afshar \cite{afshar}, generating a very heated debate in the literature \cite{debate1,debate2,debate3,debate4}. 
However, this discussion is not in the scope of the present contribution.
We first formulate the principle in a simple and operational way, relating it to the empirical unpredictability of incompatible measurements. Then it is shown that any preparation violating the principle implies the possibility of deterministically creating preparations that violate Tsirelson's bound. Moreover, these ``superquantum'' preparations could be used to distinguish and clone a plethora of non-orthogonal quantum states.

\section{Preliminaries}

\paragraph{Operational theories.} We take inspiration in the formulation of \textit{operational theories} \cite{ferrie}. 
An operational theory models mathematically a physical experiment in terms of primitive notions as preparations, measurements, outcomes and systems. 
More precisely, a preparation is a completely specified experimental procedure -- a set of mutually exclusive preparations for an experiment forms then a set $\mathcal{P}$. 
In an experiment, a preparation $P\in\mathcal{P}$ is subjected to a measurement $M$, which is an element of a set $\mathcal{M}$ of mutually exclusive measurements. 
This irreversible procedure gives some outcome $k$, which is one element of a set $\mathcal{K}$ of mutually exclusive and exhaustive outcomes. 
The goal of any operational theory is then to determine the probability $p(k|P,M)$, i.e., the probability that the outcome $k$ occurs given that we are performing the measurement $M$ of the preparation $P$. 
Shortly, an operational theory is a specification $\{\mathcal{P},\mathcal{M}, \mathcal{K},p(k|P,M)\}$. 
For example, quantum theory is an operational theory where the preparations are given by density operators, 
measurements are given by observables and the probabilities are calculated through the rule $Tr(\rho E_k)$, where $\{E_k\}$ are elements of a Positive Operator-Valued Measure (POVM) associated with the observable $M$.  

For a two-level system it is usual to work in the so-called Bloch vector representation. A preparation is fully specified by a three-dimensional Bloch vector with real components 
$\mathbf{r}=(r_x,r_y,r_z)$; for example, $x$, $y$ and $z$ are understood as orthogonal directions in space for the Stern-Gerlach apparatus and as the three independent polarization degrees 
of freedom in optical setups. 
A measurement on a two-level system  has only two outcomes, which we denote by $\pm 1$; we refer to this dichotomic measurement in direction $\mathbf{\hat{n}}$ as  $\sigma_{\mathbf{\hat{n}}}$.
It is an empirical evidence that the probabilities are calculated through the formula
\begin{eqnarray}
p(\pm 1 | \mathbf{r},\sigma_{\mathbf{\hat{n}}})=\frac{1}{2}(1\pm\mathbf{r}\cdot\mathbf{\hat{n}}). \label{blochprob}
\end{eqnarray}
The mean value of a measurement in direction $\mathbf{\hat{n}}$ is simply 
$\langle\sigma_{\mathbf{\hat{n}}}\rangle=p(+1 | \mathbf{r},\sigma_{\mathbf{\hat{n}}})-p(-1 | \mathbf{r},\sigma_{\mathbf{\hat{n}}})=\mathbf{r}\cdot\mathbf{\hat{n}}$, which can be related to Malus' law formula in classical optics and to the standard coupling between a magnetic dipole moment and an external magnetic field in classical magnetism. Indeed, this kind of representation was common place in optics before the advent of quantum theory, 
where the elements of the Bloch vector are called Stokes parameters and the Bloch sphere is also called Poincar\'e sphere. 

\paragraph{Complementarity.} Inspired by the uncertainty relations of Heisenberg \cite{up1,up2}, Bohr introduced in a series of lectures and essays \cite{bohr1,bohr2,bohr3,bohr4,bohr5,bohr6} the so-called \textit{principle of complementarity} (PC)\cite{feynman,cuffaro1,lahti1}, 
which establishes that evidences obtained under different experimental arrangements are complementary, 
in the sense that they cannot be unambiguously determined: the very meaning of acquiring information forbids us of having absolute knowledge or arbitrary precision of physical quantities for some preparations. 

To motivate the discussion, let us imagine a group of scientists that never had contact with quantum theory and receive as a gift a Stern-Gerlach apparatus and a source of spin-($1/2$) particles. 
These scientists observe that when they measure $\sigma_{\mathbf{\hat{n}}}$ in different directions, the outcomes appear with probabilities that depend on the 
preparation and on the directions that they have chosen to measure. They observe also that some special arrangements of the preparation and the direction of measurement 
yields total predictability of outcomes. For example, if they prepare the particles' beam polarized in the $z$ direction and then measure $\sigma_z$, the outcomes are 
fully determined. 
However, after trying a large number of possible different arrangements of preparations and measurements, one inevitable question will appear to them: 
``Why is it not possible to predict with certainty (probability $1$ or $0$) the outcomes of measurements in two different directions, for a fixed preparation?" 
There is in principle no rule that forbids them of obtaining, for some fixed preparation, the outcome
$+1$ with probability one in two different directions. Paraphrasing Einstein, Podolsky and Rosen \cite{epr}, it is reasonable to assume that certainty in one direction should not in any way forbid certainty in another direction. It is clear then that there is some physical law that forbids this perfectly legitimate situation. 
The basic empirical evidence is that if one measures $\sigma_{\mathbf{\hat{n}}}$ in a fixed direction $\mathbf{\hat{n}}$ and obtains $+1$ with certainty 
implies that the outcomes of measuring $\sigma_{\mathbf{\hat{m}}}$ in a different direction $\mathbf{\hat{m}}$ do not occur with total certainty. 

The discussion will be restricted mostly to two-level systems in what follows. An extension to higher dimensions is sketched in the end of the text. 
For a two-level system, the principle of complementarity reads:
\begin{principle}
For a fixed preparation, measurements of $\sigma_{\mathbf{\hat{n}}}$ and $\sigma_{\mathbf{\hat{m}}}$ in non-colinear directions 
$\mathbf{\hat{n}}$ and $\mathbf{\hat{m}}$ are not both predictable.  
\end{principle}
By \textit{predictable} we mean that the outcomes of the measurement are totally determined, i.e., one occurs with unit probability, implying the other have zero probability of occurrence. 
Thus, predictability means we can certainly know the result of measuring $\sigma_{\mathbf{\hat{n}}}$. 
Principle 1 then states that predictability of a measurement in a certain direction precludes the predictability 
of a measurement in a different direction; in this sense these different measurements are complementary. 
Let us see now how the PC imposes bounds on the Bloch vector:
\begin{observation}
For a two-level system, a preparation with Bloch vector $\mathbf{r}$ satisfies the principle of complementarity iff $r=||\mathbf{r}||\leq 1$. Equivalently, 
\begin{eqnarray}
\langle\sigma_{\mathbf{\hat{n}}_1}\rangle^2+ \langle\sigma_{\mathbf{\hat{n}}_2}\rangle^2+\langle\sigma_{\mathbf{\hat{n}}_3}\rangle^2\leq 1,\label{bound}
\end{eqnarray}
with $\mathbf{n}_1$, $\mathbf{n}_2$ and $\mathbf{n}_3$ orthogonal directions.
\end{observation}

\begin{proof}
If for a preparation with Bloch vector $\mathbf{r}$ we have $r>1$, i.e., the bound (\ref{bound}) is violated, 
writing $\mathbf{r}=r\mathbf{\mathbf{\hat{r}}}$, we have that the probability of obtaining outcome $+1$ for the measurement $\sigma_{\mathbf{\hat{n}}}$ is
\begin{eqnarray}
p(+1|\mathbf{r},\sigma_{\mathbf{\hat{n}}}) = \frac{1}{2}(1+r\mathbf{\hat{r}}\cdot\mathbf{\hat{n}}).
\end{eqnarray}
It is easy to see that there exists an infinite number of unit vectors $\mathbf{\hat{n}}$ such that $\mathbf{\hat{r}}\cdot\mathbf{\hat{n}}=1/r$ and thus $p(+1|\mathbf{r},\sigma_{\mathbf{\hat{n}}})=1$. 
Geometrically, this corresponds to the intersection between the affine plane $\mathbf{x}\cdot\mathbf{y}=1/r$ and the unit sphere $||\mathbf{\hat{x}}||=1$, which is satisfied by a circle where each
point correspond to a direction $\mathbf{\hat{n}}$ such that $p(+1|\mathbf{r},\sigma_{\mathbf{\hat{n}}})=1$. 
Thus, $\mathbf{\hat{n}}$ is fully predictable for an infinite number of non-colinear directions $\mathbf{\hat{n}}$. This proves the forward implication.
Now, for the backward implication, if the PC is violated, we have that there exist a preparation $\mathbf{\hat{r}}=r\mathbf{\hat{r}}$ and non-colinear $\mathbf{\hat{n}}$ and $\mathbf{\hat{m}}$ 
such that $\frac{1}{2}(1+r\mathbf{\hat{r}}\cdot\mathbf{\hat{n}})=\frac{1}{2}(1+r\mathbf{\hat{r}}\cdot\mathbf{\hat{m}})=1$, 
implying $r\mathbf{\hat{r}}\cdot\mathbf{\hat{n}}=r\mathbf{\hat{r}}\cdot\mathbf{\hat{m}}=1$. Thus, we must have $\mathbf{\hat{r}}\cdot\mathbf{\hat{n}}=\mathbf{\hat{r}}\cdot\mathbf{\hat{m}}=1/r$. 
Since $\mathbf{\hat{r}}$, $\mathbf{\hat{n}}$ and $\mathbf{\hat{m}}$ 
are unit vectors and non-colinear, we have that $r=(\mathbf{\hat{r}}\cdot\mathbf{\hat{n}})^{-1}=(\mathbf{\hat{r}}\cdot\mathbf{\hat{m}})^{-1}>1$, 
implying the bound (\ref{bound}) is violated and hence the backward implication and the Observation are proven. \smartqed
\end{proof}

Observation 1 implies the usual notion of complementarity for the Mach-Zender interferometer \cite{dstheo1,dstheo2,bosyk} expressed by the duality relation
(\ref{dualrel}). It is also equivalent to Larsen-Luis complementarity relations \cite{larsen,luis1} and bounds given by entropic uncertainty relations \cite{eur1,eur2,eur3}. 
Moreover, it is noteworthy that the preparations respecting the PC correspond to the Bloch ball of preparations described by quantum theory. In comparison to the original work \cite{dstheo2}, which relies on the positivity of a density matrix - and hence on the postulates of quantum theory - to reach the bound (\ref{bound}) and consequently the duality relation (\ref{dualrel}), our formulation does not fully rely on quantum theory, but only on the PC, presenting thus a slight conceptual advantage.   

It is clear that a violation of (\ref{bound}) could lead to negative or greater than one values for (\ref{blochprob}) for some directions of measurement, 
so we need to justify how to properly handle this situation. It is easy to see that the directions of measurement satisfying $|\mathbf{r}\cdot\mathbf{\hat{n}}|\leq 1$ give true values of probabilities (\ref{blochprob}).  
For directions in which $|\mathbf{r}\cdot\mathbf{\hat{n}}|> 1$, we need to specify a rule for the probabilities of the outcomes. 
There are many possible rules one could adopt -- for simplicity and for reasons that are explained later in the text, in what follows we assume that in 
directions satisfying $|\mathbf{r}\cdot\mathbf{\hat{n}}|> 1$ measurements occur randomly with maximal uncertainty, i.e., with probability $1/2$.  
This displays the form of measurement-contextuality for preparations seen in quantum theory, in the sense of \cite{spekkens3,spekkens2,blasiak2}, 
since different preparations define different sets of allowed measurements. 

\paragraph{Transformations of preparations.} Before proceeding, we need to specify how to transform one preparation into another. 
We can greatly simplify the calculations that will appear through the introduction of the well-known operator
\begin{eqnarray}
\rho(\mathbf{r}) =\frac{1}{2}(I+\mathbf{r}\cdot\mathbf{\sigma}),\label{blochball} 
\end{eqnarray}
which in optics is called the polarization matrix, where $\mathbf{r}$ is the Bloch vector associated to the preparation and $\mathbf{\sigma}$ is a vector composed by the Pauli matrices. This operator is hermitean and unit-trace; 
as shown in Observation 1, 
a preparation respects the PC iff $r\leq 1$, which means that $\rho(\mathbf{r})$ is a positive operator and corresponds to a density matrix. If we identify a measurement 
$\sigma_{\mathbf{\hat{n}}}$ with the operator $\mathbf{\sigma}\cdot\mathbf{\hat{n}}$ then we can use the mathematical machinery of quantum operators to simplify our discussion. 
In order to see this, we define the projector $\Pi_{\mathbf{\hat{n}}}=(1/2)(I+\mathbf{\hat{n}}\cdot\mathbf{\sigma})$; in the optics literature these are known as 
Jones' matrices. It is trivial that $\Pi_{\mathbf{\hat{n}}}+\Pi_{-\mathbf{\hat{n}}}=I$ and $\mathbf{\sigma}\cdot\mathbf{\hat{n}}=\Pi_{\mathbf{\hat{n}}}-\Pi_{-\mathbf{\hat{n}}}$. 
The rule (\ref{blochprob}) can then be rewritten as $p(\pm 1 | \mathbf{r},\sigma_{\mathbf{\hat{n}}})=Tr[\Pi_{\mathbf{\pm\hat{n}}}\rho(\mathbf{r})]$.
We will assume then that the set of allowed transformations are composed by standard completely-positive trace-preserving linear maps over the set of operators $\rho(\mathbf{r})$. 
An operation over $\rho(\mathbf{r})$ will then induce a change over $\mathbf{r}$ corresponding to usual processes in a two-level system experiment. 
For example, local unitaries over $\rho(\mathbf{r})$ correspond to rotations of the Bloch vector $\mathbf{r}$ and 
the matrix $P(\mathbf{r}, \mathbf{r}')=\rho(\mathbf{r})\otimes\rho'(\mathbf{r}')$ represents the addition of an extra two-level system with Bloch vector $\mathbf{r}'$,
where $\otimes$ is the kronecker product of the individual matrices. We are assuming, as well, the validity of L\"uders' rule for the description of a preparation after the occurrence of a measurement.  
Similar translations of multipartite two-level systems operations in terms of Bloch vector operations can be found in the literature \cite{altafini1,altafini2}.
Thus, for preparations respecting the PC, there is no deviation from standard predictions of quantum theory. 
Our formulation can be seen in this equivalent way as an extension of quantum theory in order to consistently account 
for a violation of complementarity. More precisely, the rule $p(\pm 1 | \mathbf{r},\sigma_{\mathbf{\hat{n}}})=Tr[\Pi_{\mathbf{\pm\hat{n}}}\rho(\mathbf{r})]$ by itself does not 
rules out negative operators $\rho(\mathbf{r})$, since for many directions of measurements, $\sigma_{\mathbf{\hat{n}}}$, the values $p(\pm 1 | \mathbf{r},\sigma_{\mathbf{\hat{n}}})$ 
are genuine probabilities -- and this is enough to make predictions about the system at hand. 
For two-level systems, Observation 1 shows that the PC is equivalent to imposing the postulate of positive operators $\rho(\mathbf{r})$ \cite{lahti2}, 
while a violation of that principle would demand the abdication of this postulate and a
legitimate use of negative operators to represent preparations. 
We observe that similar extensions of quantum theory in terms of non-positive operators have been employed to represent nonlocal boxes \cite{acin,alsafi}, for the construction of efficient simulation
schemes \cite{brukner}, toy models of quantum theory \cite{spekkens1,blasiak} and more recently to locally extend quantum mechanics in the formulation known as ``boxworld" \cite{boxworld,brunner}. 

\section{Results}

\paragraph{Nonlocal Box creation.} We restrict our discussion to the standard scenario where two observers perform dichotomic measurements 
$A_1$ and $A_2$ (first observer) and $B_1$ and $B_2$ (second observer). Defining the Bell operator
\begin{eqnarray}
\mathcal{B}= A_1B_1 + A_1B_2 + A_2B_1 - A_2B_2,
\end{eqnarray}
it is well-known that assumptions of locality, realism and free-choice impose the Clauser-Horne-Shimony-Holt (CHSH) bound \cite{chsh} $|\langle\mathcal{B}\rangle|\leq 2$. 
The maximal violation attainable by quantum states and measurements is the so-called \textit{Tsirelson's bound} $|\langle\mathcal{B}\rangle|\leq 2\sqrt{2}$ \cite{tsirelson}. 
As shown by Popescu and Rohrlich \cite{pr}, there are non-signalling probability distributions which violate Tsirelson's bound and some even reach the maximum algebraic value 
$|\langle\mathcal{B}\rangle| = 4$. These theoretical constructions can be studied in the framework of nonlocal boxes \cite{nlbox}. 
We now employ these ideas and constructions to show that violations of complementarity allow the construction of nonlocal boxes violating Tsirelson's bound.
\begin{theorem}
For a two-level system, any preparation violating the principle of complementarity enables the deterministic generation of a bipartite preparation that violates Tsirelson's bound. 
\end{theorem}
\begin{proof}
According to Observation 1, a preparation with Bloch vector $\mathbf{r}$ violates the complementarity principle iff $r>1$. Without loss of generality, let us consider that $\mathbf{\hat{r}}=\mathbf{\hat{x}}=(1,0,0)$. Using the equivalent representation given the matrix (\ref{blochball}), we add an ancillary quantum state $\rho'=(1/2)(I+\sigma_x)$ and thus resulting with the extended state
\begin{eqnarray}
\rho(\mathbf{r})\otimes\rho'=\frac{1}{4}(I+r\sigma_x)\otimes(I+\sigma_x).
\end{eqnarray}
The measurements of the observable $\sigma_z\otimes\sigma_z$ on this preparation yields results $\pm 1$ with probability $1/2$ each. If outcome $+1$ occurs, nothing else is done; if outcome $-1$ occurs, the unitary $\sigma_x$ is applied on the second subsystem. Hence, the preparation 
\begin{eqnarray}
P=\frac{1}{2}[(1+r)P_{Bell+}+(1-r)P_{Bell-}],\label{nlbox2}
\end{eqnarray}
is deterministically generated, where 
\begin{eqnarray}
P_{Bell\pm}=\frac{1}{2}(I\otimes I \pm\sigma_x\otimes\sigma_x\mp\sigma_y\otimes\sigma_y+\sigma_z\otimes\sigma_z).
\end{eqnarray} 
For $r\leq\sqrt{2}$, let us choose $A_1=(\sigma_x+\sigma_y)/\sqrt{2}$, $A_2=(\sigma_x -\sigma_y)/\sqrt{2}$ $B_1=\sigma_x$ and 
$B_2=-\sigma_y$. It is easy to see that for these local measurements we have $\langle\mathcal{B}\rangle=Tr(\mathcal{B}P) = 2\sqrt{2}r$, i.e., a violation of Tsirelson's bound whenever $1<r\leq\sqrt{2}$. 
For $r>\sqrt{2}$, we choose $A_1=(\sigma_x+\sigma_y)/\sqrt{2}$, $A_2=(\sigma_x -\sigma_y)/\sqrt{2}$ $B_1=(\frac{\sqrt{2}}{r})\sigma_x+(\frac{\sqrt{r^2-2}}{r})\sigma_y$ and 
$B_2=(\frac{\sqrt{r^2-2}}{r})\sigma_y-(\frac{\sqrt{2}}{r})\sigma_y$, obtaining $\langle\mathcal{B}\rangle=Tr(\mathcal{B}P)=4$ for any value of $r$, i.e., the maximal violation of Tsirelson's bound that 
does not violate non-signalling. \smartqed
\end{proof}
The preparation (\ref{nlbox2}) was originally proposed in \cite{acin} in order to represent post-quantum nonlocal boxes and we have shown explicitly the set of measurements that enables the violation of Tsirelson's bound $|\langle\mathcal{B}\rangle|\leq 2\sqrt{2}$ whenever $r>1$. Hence, there is a 
deep link between bounds imposed locally by complementarity and bounds on nonlocal correlations.

\paragraph{Distinguishability and cloning.}

The preparations we are considering, even those violating the PC, respect linearity and thus it is expected that some kind of no-go theorem for distinguishability and cloning is still valid in a more general sense. We first establish a condition for a pair of preparations to be jointly-clonable, i.e., a condition on these states such that they can be cloned by the same deterministic process. 
\begin{theorem}
If two preparations with Bloch vectors $\mathbf{r}$ and $\mathbf{r}'$ are jointly-clonable then $\mathbf{r}\cdot\mathbf{r}'=\pm 1$,
\end{theorem}
\begin{proof}
Let the preparations with Bloch vectors $\mathbf{r}$ and $\mathbf{r}'$ be joint-clonable. 
Using (\ref{blochball}), these preparations correspond to matrices $\rho(\mathbf{r})$ and $\rho'(\mathbf{r}')$. 
If these preparations are joint-clonable, then there exists an unitary $U$ such that
\begin{eqnarray}
U(\rho\otimes |e_0\rangle\langle e_0|) U^{\dagger} = \rho\otimes\rho; \ \ \ \ U(\rho'\otimes |e_0\rangle\langle e_0|) U^{\dagger} = \rho'\otimes\rho',
\end{eqnarray}
where $|e_0\rangle$ is a fixed pure state. We then have
\begin{eqnarray}
Tr[(\rho\otimes\rho)(\rho'\otimes\rho')]= Tr[U(\rho\otimes |e_0\rangle\langle e_0|)U^{\dagger}U(\rho'\otimes |e_0\rangle\langle e_0|)U^{\dagger}] = Tr(\rho\rho'),\nonumber\\
&&
\end{eqnarray}
where we used the cyclic property of the trace in the last step. Since $Tr(A\otimes B)=Tr(A)Tr(B)$, the first term is equal to $[Tr(\rho\rho')]^2$. Thus we have
\begin{eqnarray}
[Tr(\rho\rho')]^2 = Tr(\rho\rho'),
\end{eqnarray}
as a condition to existence of a unitary $U$ that clones $\rho$ and $\rho'$. This is equivalent to $Tr(\rho\rho')=0$ or $Tr(\rho\rho')=1$, 
which is equivalent to $\mathbf{r}\cdot\mathbf{r}'=\pm 1$. \smartqed
\end{proof} 

The equations $\mathbf{r}\cdot\mathbf{r}'=\pm 1$  are those of two affine planes that cross the interior of the Bloch ball, 
whenever at least one of the preparations $\mathbf{r}$ or $\mathbf{r}'$ violates the PC. 
Remarkably, still there are  states that are not able to be jointly distinguished/cloned, suggesting fundamental limits even in the case of strong violations of physical principles. 

\begin{figure}[h]\centering\includegraphics[scale=0.5]{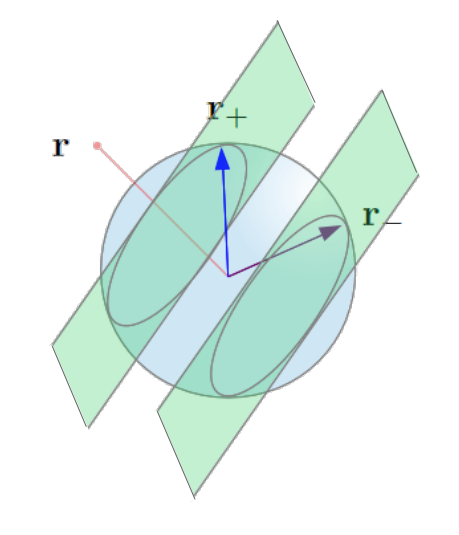} \label{planes}\caption{(Color Online) A preparation $\mathbf{r}$ violating the PC (in red) defines two planes (in green) crossing the Bloch sphere (in light blue). 
These planes are formed by the preparations whose Bloch vectors satisfy $\mathbf{r}\cdot\mathbf{r_{\pm}}=\pm 1$ .}\end{figure}

The following result shows that an arbitrary preparation violating the PC can be used to intermediate the distinguishability of some non-orthogonal quantum states. This enables naturally a protocol to clone these two states.
\begin{theorem}
Given a preparation with Bloch vector $\mathbf{r}$ violating the principle of complementarity, 
the quantum states with Bloch vectors $\mathbf{r}_{\pm}$ satisfying $\mathbf{r}\cdot\mathbf{r_{\pm}}=\pm 1$ are distinguishable by a deterministic protocol.
\end{theorem}
The main idea to prove this result is to design a measurement where each outcome corresponds to a perfect correlation between $\mathbf{r}$ and $\mathbf{r}_+$ and $\mathbf{r}$ and $\mathbf{r}_-$ exclusively.
\begin{proof}
Using (\ref{blochball}), a preparation with Bloch vector $\mathbf{r}$ is represented by the matrix 
\begin{eqnarray}
\rho(\mathbf{r})= \frac{1}{2}(I+r\mathbf{\sigma}\cdot\mathbf{\hat{r}}),
\end{eqnarray}
Let $\mathbf{r}_+$ and $\mathbf{r}_-$ be the Bloch vectors of two quantum states such that $\mathbf{r}\cdot\mathbf{r}_{\pm}=\pm 1$.
Let $\rho_{\pm}$ be the matrices (\ref{blochball}) representing the preparations with Bloch vectors $\mathbf{r}_{\pm}$; then it is straightforward 
that $Tr(\rho\rho_-)=0$, $Tr(\rho\rho_+)=1$. These quantum states are expressed as 
\begin{eqnarray}
\rho_{\pm} = \frac{1}{2}\left[I\pm\frac{1}{r}(\mathbf{\sigma}\cdot\mathbf{\hat{r}})+y(\mathbf{\sigma}\cdot\mathbf{\hat{m}})+z(\mathbf{\sigma}\cdot\mathbf{\hat{n}})\right], 
\end{eqnarray}
where $\mathbf{\hat{m}}$ and $\mathbf{\hat{n}}$ are directions orthogonal to $\mathbf{\hat{r}}$ and the real numbers $y$ and $z$ satisfy $(1/r^2)+y^2+z^2\leq 1$. Without loss of generality, 
we assume that $\mathbf{\hat{r}}$, $\mathbf{\hat{m}}$ and $\mathbf{\hat{n}}$ form a right-hand triple of vectors. 
We see that the projectors on Bell states are written as
\begin{eqnarray*}
\Pi_{\phi_{\pm}} =\frac{1}{4}[I^{\otimes 2}\pm(\mathbf{\sigma}\cdot\mathbf{\hat{r}})^{\otimes 2}\mp(\mathbf{\sigma}\cdot\mathbf{\hat{m}})^{\otimes 2} +(\mathbf{\sigma}\cdot\mathbf{\hat{n}})^{\otimes 2}], \\ \Pi_{\psi_{\pm}} =\frac{1}{4}[I^{\otimes 2}\pm(\mathbf{\sigma}\cdot\mathbf{\hat{r}})^{\otimes 2}\pm(\mathbf{\sigma}\cdot\mathbf{\hat{m}})^{\otimes 2} -(\mathbf{\sigma}\cdot\mathbf{\hat{n}})^{\otimes 2}].
\end{eqnarray*}
Considering the probabilities of measurements on Bell basis for the states $\rho\otimes\rho_{\pm}$, we see that 
\begin{eqnarray}
Tr[\Pi_{\phi_+}(\rho\otimes\rho_-)] = 0, \ \ \ \  Tr[\Pi_{\phi_+}(\rho\otimes\rho_+)] = 1/2, \\
Tr[\Pi_{\psi_+}(\rho\otimes\rho_-)] = 0, \ \ \ \ Tr[\Pi_{\psi_+}(\rho\otimes\rho_+)] = 1/2, \\
Tr[\Pi_{\phi_-}(\rho\otimes\rho_-)] = 1/2, \ \ \ \ Tr[\Pi_{\phi_-}(\rho\otimes\rho_+)] = 0, \\
Tr[\Pi_{\psi_-}(\rho\otimes\rho_-)] = 1/2, \ \ Tr[\Pi_{\psi_-}(\rho\otimes\rho_+)] = 0.
\end{eqnarray}
Defining the projectors $\Pi_{\pm}=\Pi_{\phi_{\pm}}+\Pi_{\psi_{\pm}}$ and given that $\Pi_++\Pi_-=I$, we define a two-outcome observable $M=\Pi_+-\Pi_-$. 
From the results above it is straightforward that $Tr(\Pi_+\rho\otimes\rho_+)=1$ and $Tr(\Pi_-\rho\otimes\rho_-)=1$.
Thus, if we have a state $\mu$ that is either $\rho_+$ or $\rho_-$ - but  we do not know which - we measure $M$ on the state $\rho\otimes\mu$, 
obtaining with unit probability the outcome $\pm 1$ iff $\mu$ corresponds to $\rho_{\pm}$.\smartqed
\end{proof}
The preparation $\rho\otimes\mu$ will be undisturbed by the measurement $M$, 
since 
the output from the protocol of Theorem 3 will be deterministically either $\rho\otimes\rho_+$ or $\rho\otimes\rho_-$. 
Since for an arbitrary state there is always a non-universal deterministic cloning protocol \cite{nielsen}, 
after discriminating which state $\mu$ is we have the following result:
\begin{corollary}
Given a preparation with Bloch vector $\mathbf{r}$ violating the principle of complementarity, 
the quantum states with Bloch vectors $\mathbf{r}_{\pm}$ satisfying $\mathbf{r}\cdot\mathbf{r_{\pm}}=\pm 1$ are clonable by a deterministic protocol.
\end{corollary}
For completeness, we refer the reader to other approaches to clone non-orthogonal states using closed time-like curves \cite{ctc1,ctc2}, which however rely on some form of nonlinear dynamics.

As previously stated, for a given preparation $\rho(\mathbf{r})$ and directions of measurement $\mathbf{\hat{n}}$ such that $|\mathbf{r}\cdot\mathbf{\hat{n}}|>1$, we ascribe the rule that the outcomes occur randomly. Besides its mathematical simplicity, we justify this choice based on a reasoning similar to the proofs of Theorem 2 and Theorem 3. Let us then consider a different rule, one in which the outcome $+1$ of measuring $\mathbf{\sigma}\cdot\mathbf{\hat{n}}$ on $\rho(\mathbf{r})$ never occurs. From the axioms of probability  $p(+1|\mathbf{r},\mathbf{\hat{n}})+p(-1|\mathbf{r},\mathbf{\hat{n}})=1$ and the outcome $-1$ occurs with unit probability. Define the pure quantum states
$\rho(\mathbf{\hat{n}}) = (1/2)(I+\mathbf{\hat{n}}\cdot\sigma)$ and $\rho(\mathbf{\hat{n}_+}) = (1/2)(I+\mathbf{\hat{n}_+}\cdot\sigma)$, 
where $\mathbf{\hat{n}}_+$ is a unit vector such that $\mathbf{r}\cdot\mathbf{\hat{n}}_+=1$. Thus $\rho(\mathbf{\hat{n}}_+)$ is a pure quantum state on the upper circle in Fig. 1, the intersection between the plane of states that satisfy $\mathbf{r}\cdot\mathbf{r}_+=1$ and the Bloch sphere. It is possible then to distinguish the pure quantum states $\rho(\mathbf{\hat{n}_+})$ and $\rho(\mathbf{\hat{n}})$ in a simple fashion: if we measure $(\sigma\cdot\mathbf{\hat{n}})\otimes(\sigma\cdot\mathbf{\hat{n}})$ on $\rho(\mathbf{r})\otimes\rho(\mathbf{\hat{n}})$, the outcome $-1$ occurs with unit probability, while if we measure $(\sigma\cdot\mathbf{\hat{n}}_+)\otimes(\sigma\cdot\mathbf{\hat{n}}_+)$ on $\rho(\mathbf{r})\otimes\rho(\mathbf{\hat{n}_+})$, the outcome $+1$ occurs with unit probability. 

Hence, one could discriminate between \textit{any} pure quantum state $\rho(\mathbf{\hat{n}})$ and the pure quantum states $\rho(\mathbf{\hat{n}}_+)$ on the upper circle of Fig.1.  The pure quantum states $\rho(\mathbf{\hat{n}})$ correspond geometrically to sections of the Bloch sphere, whose area is non-negligible compared to the null-measure area of the circle formed by states $\rho(\mathbf{\hat{n}}_+)$. 
For increasing values of $r$, the planes in Fig.1 get closer in distance and the area formed by the set of states $\rho(\mathbf{\hat{n}})$ increases. For $r\rightarrow\infty$, this area tends to the area of the whole Bloch sphere and one could then distinguish the whole Bloch sphere\footnote{Minus the null measure set constituted by the states $\rho(\mathbf{\hat{n}}_+)$ on the equatorial circle of the Bloch sphere.} from the preparation $\rho(\mathbf{\hat{n}})\otimes\rho(\mathbf{\hat{n}_+})$; an arbitrary pure state in the Bloch sphere could then be jointly-cloned with $\rho(\mathbf{\hat{n}})\otimes\rho(\mathbf{\hat{n}_+})$, which would be a clear violation of the No-Cloning Theorem. This extreme example illustrates that certainty in the region $|\mathbf{r}\cdot\mathbf{\hat{n}}|>1$ could imply strong violations of No-Cloning. In order to ensure that no distinguishability/cloning advantage would come from measurements such that $|\mathbf{r}\cdot\mathbf{\hat{n}}|>1$, we impose random outcomes in this situation, but we leave as an open question whether less stringent assignments of probabilities rules could guarantee consistent results concerning distinguishability and cloning.

\section{Higher dimensions}

For two-level systems it was shown that violations of the PC implies the possibility of violating Tsirelson's bound and in breaking the limits of distinguishability and cloning protocols. 
The distinctive feature in this situation was the relative independence on the typical rules associated to quantum theory, through the Bloch vector representation. 
Neverthless, through (\ref{blochball}) we argued that our formulation is equivalent to an extension of quantum theory 
in terms of non-positive operators for the preparations. 
We adopt this approach in order to formulate complementarity for higher-dimensional systems, i.e., we introduce an extension of the quantum theory that 
does not impose positive-semidefiniteness on the operators representing preparations. 

Explicitly, our ``toy model'' has the set of preparations $\mathcal{P}$ composed of self-adjoint unit-trace operators $\tilde{\rho}$, the set of measurements $\mathcal{M}$ composed of 
self-adjoint operators $M$ and the probabilities are calculated via the trace-rule $p(k|\tilde{\rho},M)=Tr(\tilde{\rho}E_k)$ where $\{E_k\}$ is the POVM associated to $M$; 
it is noteworthy that within Hilbert-space formulations the trace-rule is unique \cite{gleason}. 
Once again we consider only the results of measurements that give genuine values of probabilities $p(k|\tilde{\rho},M)$.
Let us introduce some definitions:
\begin{definition}
Two non-degenerate measurements $M$ and $N$ are fully incompatible if they do not share any eigenstate. 
\end{definition}
This definition captures the intuitive notion that a measurement is always disturbed if it is followed by a measurement that is fully incompatible with it, independent on the preparation that is measured. 
\begin{definition}
The outcomes of a non-degenerate observable are predictable if one of them occurs with unit probability.  
\end{definition}
By \textit{predictable} we mean that the outcomes of the measurement are totally determined, i.e., one occurs with unit probability, implying the others have zero probability of occurrence. 
Then we state our version of the PC for higher-dimensional systems:
\begin{principle}
Given a fixed preparation, the outcomes of measurements of two non-degenerate fully incompatible measurements are not both predictable.
\end{principle}
This principle expresses the complementary aspect of fully incompatible measurements, since predictability of one quantity implies 
unpredictability of another quantity that is fully incompatible with it. It is easy to see that Principle 1 is a special case 
of Principle 2, when one uses the equivalent representation (\ref{blochball}) and the identification $\sigma_{\mathbf{\hat{n}}}\equiv\mathbf{\sigma}\cdot{\mathbf{\hat{n}}}$.
Mathematically, given $\Pi_{\nu}=|\nu\rangle\langle\nu|$, the principle simply states that there is no preparation $\tilde{\rho}$ 
for which $Tr(\tilde{\rho} \Pi_{\psi})=Tr(\tilde{\rho} \Pi_{\phi})=1$, when $\phi$ and $\psi$ are non-orthogonal. 

Observation 1 shows that a two-level system preparation satisfies the PC iff the operator representing it is positive semidefinite. 
If this equivalence would hold as well for arbitrary dimensions, 
then complementarity would be the principle explaining the quantum bounds on non-local correlations.
For higher-dimensional systems, however, this is not the case and the principle does not rule out all negative operators. 
To illustrate the main problems, let us consider a three-level system with orthonormal basis 
$\{|b_0\rangle,|b_1\rangle,|b_2\rangle\}$. The operator $\eta=(0.85)|b_0\rangle\langle b_0|+(0.25)|b_1\rangle\langle b_1|-(0.1)|b_2\rangle\langle b_2|$ 
is an example of non-positive operator that satisfies the PC, since its maximal eigenvalue is smaller than $1$ and there is no rank-$1$ projective measurement 
for which the probability $\langle\psi|M|\psi\rangle$ is unit - the maximal eigenvalue of an operator is the maximal value of $\langle\psi|M|\psi\rangle$ \cite{horn}. 
Hence, for higher-dimensional systems violation of the PC does not rule out completely preparations beyond quantum mechanics. 
Interestingly, preparations that do violate the PC as formulated here are still able to enhance the tasks of distinguishability and cloning in the same lines as the two-level case; 
the full argument is shown in the next section. Hence, violation of the PC is at least a necessary condition for performing beyond-quantum tasks. 

A possible solution is to see the limitations imposed by complementarity as one in a series of conditions that a positive operator must fulfill. For a $N$-dimensional system, consider the following symmetric functions of the eigenvalues $\lambda_0, \lambda_1,\ldots,\lambda_{N-1}$ of an operator $\rho$:
\begin{eqnarray}
s_1=\sum_i \lambda_i, \ \ \ s_2=\sum_{i<j} \lambda_i\lambda_j, \ \ \ s_3=\sum_{i<j<k} \lambda_i\lambda_j\lambda_k, \ \ \ \ldots \,.
\end{eqnarray}
These functions are related \cite{zyc} to the moments $Tr(\rho^k)$ of the operator $\rho$ through the recursive formulae:
\begin{eqnarray*}
s_1 = Tr(\rho), \\ s_2 = \frac{1}{2}[s_1Tr(\rho)-Tr(\rho^2)], \\ \vdots \\ s_k=\frac{1}{k}[s_{k-1}Tr(\rho)-s_{k-2}Tr(\rho^2)+\ldots+(-1)^{k-1}Tr(\rho^k)], \\ \vdots  
\end{eqnarray*}
The operator $\rho$ is then positive-semidefinite iff $s_k\geq 0$, $k=1,2,\ldots,N$ and inverting the above relations in terms of the moments $Tr(\rho^k)$ gives a series of conditions on $\rho$. The first condition corresponds to the well-known normalization, $s_1=Tr(\rho)=1$. The second condition is $Tr(\rho^2)\leq 1$ and can be identified with the PC, since in any Bloch vector representation it corresponds to $||\mathbf{r}||^2\leq g$, where $g$ is a positive value dependent on the particular Bloch vector representation used -- for two-level systems it gives the relation (\ref{bound}). For the $3$-level operator $\eta=(0.85)|b_0\rangle\langle b_0|+(0.25)|b_1\rangle\langle b_1|-(0.1)|b_2\rangle\langle b_2|$, we see that that conditions $Tr(\eta)=1$ and $Tr(\eta^2)\leq 1$ are satisfied, but the third condition $Tr(\eta^3)-(3/2)Tr(\eta^2)+(1/2)\geq 0$ is violated, since $Tr(\eta^3)-(3/2)Tr(\eta^2)+(1/2)=-0.06375$. For higher-dimensional systems, higher-order tests based on the various moments single out the set of positive operators of quantum theory; the physical interpretation of such tests, however, are not so clear as complementarity $Tr(\rho^2)\leq 1$. Moreover, for $N$-level systems there are different Bloch ball representations of an operator\cite{blochqudits1,blochqudits2}, each suited to a specific experimental situation and mathematically the geometrical problems that arise are quite challenging and beyond the scope of the present work.     

\paragraph{Distinguishability in higher dimensions.} As explained previously, in order to violate the PC it is necessary that the operator $\rho$ representing the preparation has at least one eigenvalue bigger than $1$. 
Thus, an arbitrary preparation violating the PC in spectral decomposition reads 
\begin{eqnarray}
\tilde{\rho}=(1+\epsilon)|\psi_0\rangle\langle\psi_0|+\sum_{k=1}^{d-1}\lambda_k |\psi_k\rangle\langle\psi_k|,
\end{eqnarray}
where the $\{\psi_n\}$ ($\{\lambda_n\}$) are the eigenvectors (eigenvalues) of $\tilde{\rho}$, $\epsilon$ is a positive real number and $\epsilon +\sum_{k=1}^{d-1}\lambda_k=0$, implying $Tr\rho=1$. 
Define an arbitrary pure state $|\nu_k\rangle=\sum_n\alpha^{(k)}_n|\psi_n\rangle$, with $\sum_n|\alpha^{(k)}_n|^2=1$. Then we have 
\begin{eqnarray}
\langle\nu_i|\tilde{\rho}|\nu_i\rangle = (1+\epsilon)|\alpha^{(i)}_0|^2+\sum_k\lambda_k|\alpha^{(i)}_k|^2. 
\end{eqnarray}
For simplicity, let us consider first a vector $|\nu_1\rangle$, such that $|\alpha^{(1)}_1|^2=|\alpha^{(1)}_2|^2=\ldots=|\alpha^{(1)}_{d-1}|^2$, implying $|\alpha^{(1)}_0|^2+(d-1)|\alpha^{(1)}_1|^2=1$. 
Then $\langle\nu_1|\rho|\nu_1\rangle=1$ is equivalent to
\begin{eqnarray}
(1+\epsilon)|\alpha^{(1)}_0|^2+(\sum_{k=1}^{d-1}\lambda_k)|\alpha^{(1)}_1|^2 = 1 \Rightarrow (1+\epsilon)|\alpha^{(1)}_0|^2-\epsilon|\alpha^{(1)}_1|^2 = 1,
\end{eqnarray}
where we used the relation $\epsilon +\sum_{k=1}^{d-1}\lambda_k=0$. Since $|\alpha^{(1)}_0|^2+(d-1)|\alpha^{(1)}_1|^2=1$, one easily finds the solution
\begin{eqnarray}
|\alpha^{(1)}_0|^2 = \frac{\epsilon+d-1}{d\epsilon + d-1}, 
\end{eqnarray}
and then an infinite number of vectors $|\nu_1\rangle$ such that $\langle\nu_1|\tilde{\rho}|\nu_1\rangle=1$, i.e., such that the PC is violated by $\tilde{\rho}$. 
By the same reasoning, defining a vector 
$|\nu_0\rangle$ such that $|\alpha^{(0)}_1|^2=|\alpha^{(0)}_2|^2=\ldots=|\alpha^{(0)}_{d-1}|^2$ but such that $\langle\nu_0|\tilde{\rho}|\nu_0\rangle=0$ gives the solution
\begin{eqnarray}
|\alpha^{(0)}_0|^2 = \frac{\epsilon}{d\epsilon + d-1}, 
\end{eqnarray}
which is fullfilled by an infinite number of vectors as well.

Let us design then a POVM discriminating quantum states in the form $|\nu_0\rangle$ from those in the form $|\nu_1\rangle$. Define the following maximally entangled states
\begin{eqnarray}
|\phi_k\rangle =\frac{1}{\sqrt{d}}\sum_{j=0}^{d-1}\omega_d^{jk}|\psi_k,\psi_k\rangle
\end{eqnarray}
where $\omega_d=e^{i(2\pi/d)}$ is the $d$-th rooth of unity; define the projector $\Pi_1=\sum_{k=0}^{d-1}|\phi_k\rangle\langle\phi_k|$. A straightforward calculation shows that 
$Tr(\Pi_1\tilde{\rho}\otimes|\nu_1\rangle\langle\nu_1|)=1$ and $Tr(\Pi_1\tilde{\rho}\otimes|\nu_0\rangle\langle\nu_0|)=0$. Thus, defining $\Pi_0=I-\Pi_1$, we have a POVM $\{\Pi_0,\Pi_1\}$ such that 
$Tr(\Pi_1\tilde{\rho}\otimes|\nu_1\rangle\langle\nu_1|)=1$ and $Tr(\Pi_0\tilde{\rho}\otimes|\nu_0\rangle\langle\nu_0|)=1$, i.e., we can discriminate with certainty $|\nu_0\rangle$ from 
the (almost always) non-orthogonal $|\nu_1\rangle$ and similarly to Corollary 1, clone these states deterministically as well.

\section{Conclusions}

In this work, we gave a simple and operational formulation of the principle of complementarity in terms of the empirical unpredictability of fully incompatible measurements. 
For two-level systems it was shown that a violation of complementarity is equivalent to: (i) the creation of nonlocal preparations that violate Tsirelson's bound, without violating non-signalling in the framework of the CHSH inequality, by using solely deterministic operations; (ii) distinguishability and hence cloning of a plethora of non-orthogonal quantum states via deterministic protocols. 
Theorem 1 seems to contradict the results of \cite{gross}, which prove that all reversible dynamics are trivial in the boxworld representation of non-signalling correlations. 
We stress that our formulation does not necessarily satisfies all rules of the boxworld formalism, having more freedom in handling preparations. We believe this is the main reason for the 
discrepancy in results.

For higher-dimensional systems the equivalence between preparations satisfying complementarity and the set of quantum states does not hold completely, but violations of complementarity were shown as necessary for the enhancement of distinguishability and cloning. 
Thus, one can see our results as giving even stronger reasons for complementarity as a major physical principle and we believe it is, if not the main reason, one 
strong argument ruling out beyond-quantum phenomema in nature. 
Moreover, we supplemented the mathematical bounds given by complementarity with higher-order bounds on the moments of a preparation. The physical interpretation of these higher-order tests is an interesting open problem whose elucidation could clarify the characterization of correlations displayed by quantum systems. 

\section*{Acknowledgments}
We are thankfull to C. Brukner, E. F. Galv\~ao, M. F. Cornelio, A. V. Saa, R. A. Mosna and P. E. M. F. Mendon\c ca for comments and suggestions in previous versions of this work and to P. Blasiak, C. Budroni and M. Kleinmman for comments and discussions that helped to improve the ideas and the overall presentation of this work. 
We are thankfull as well to M. Huber for important suggestions regarding higher-dimensional systems. 
This work is part of the institutional project $346/2017$ from the Universidade Federal de Mato Grosso and was supported by CAPES, by the EU (Marie Curie CIG 293993/ENFOQI), the BMBF (Chist-Era Project QUASAR), the FQXi Fund (Silicon Valley Community Foundation), and the DFG.
MCO acknowledges support from CNPq/FAPESP through the Instituto Nacional de Ci\^encia e Tecnologia em Informa\c c\~ao Qu\^antica (INCT-IQ) and FAPESP through the Research
Center in Optics and Photonics (CePOF).

\end{document}